\documentclass[a4paper,11pt]{article}
\pdfoutput=1 

\usepackage{jinstpub} 

\title{\boldmath A Thin Foil-foil Proton Recoil Spectrometer for DT neutrons using annular silicon detectors}

\author[1]{B. Marcinkevicius \note{Corresponding author.},}
\author{E. Andersson Sund\'en,}
\author{S. Conroy, }
\author{G. Ericsson, }
\author{A. Hjalmarsson}

\affiliation{Uppsala University, Uppsala, Sweden}

\emailAdd{benjaminas.marcinkevicius@physics.uu.se}

\abstract{	
	The use of Thin-foil proton recoil (TPR) spectrometers  to measure neutrons from Deuterium-Tritium (DT) fusion plasma has been studied previously  and is a well established technique for neutron spectrometry. The study presented here focuses on the optimisation of the TPR spectrometer configurations consisting of $\Delta $E and E silicon  detectors. In addition  an investigation of the spectrometer's ability to determine fuel ion temperature and fuel ion density ratio in ITER like DT  plasmas has been performed.
	
	A Python code was developed for the purpose of calculating detection efficiency and energy resolution as a function of several spectrometer geometrical parameters. An optimisation of detection  efficiency for selected values of resolution was performed regarding the geometrical spectrometer parameters
	using a multi-objective optimisation, a.k.a Pareto plot analysis. Moreover, the influence of detector segmentation on spectrometer energy resolution and efficiency was investigated. The code also produced response functions for the two selected spectrometer configurations. The SPEC code was used to simulate  the spectrometer's performance in determining the fuel ion temperature and  fuel ion density ratio  $n_t$/$n_d$.
	
	The results presented include the selected spectrometer configuration with calculated energy resolution and efficiency. For a selected spectrometer resolution of 5\% a maximum efficiency  of around 0.003\% was achieved. Moreover, the detector segmentation allows for a 20\% increase in spectrometer efficiency for an energy resolution of  4.3\%. The ITER requirements for a 20\% accuracy on the $n_t$/$n_d$ ratio determination and 10\% on the temperature  determination within a 100~ms sampling window can be achieved using a combination of several TPR's of same type, in order to boost efficiency.
}


\arxivnumber{} 

\collaboration[]{}

\begin{document}
	\maketitle
	\flushbottom
	
	\section{Introduction}
	\label{sec:intro}
	
	Thermal Deuterium-Tritium (DT) fusion plasmas produce neutrons with a most probable energy of 14~MeV. The neutrons escape the magnetically confined plasma and their energy distribution can be used to determine plasma parameters like ion temperature, fusion power, fuel density  as well as fuel ion ratio~\cite{Ntech,Fisher1983,Henrik2010,Marocco2012,Ericsson2010,Hellesen2012,Hellesen2015}. One of the possible methods for neutron spectroscopy is the Thin-Foil Proton Recoil (TPR)  technique which can achieve good energy resolution and has the potential for improved background separation \cite{Conroy2008,Cazzaniga2015,hawkes1999}. Recently the TPR spectrometer has been put forward as one of the techniques applied in the High Resolution Neutron Spectrometer system for ITER. A review of the spectrometers under consideration can be found in \cite{Sunden2013}. This paper focuses on two topics: simulation of the TPR spectrometer configuration and application of the spectrometer in different plasma scenarios for extraction of physics information.
	
	For the first topic, a Monte-Carlo  based code PROTON was developed to determine the efficiency and energy resolution of the TPR spectrometer  for different spectrometer configurations. The results were then used to find \cite{2004Optimization} the best efficiency for a given resolution. This lead to two different near-optimal configurations  corresponding to the use of non-segmented and segmented detectors.

	For these two configurations, spectrometer response functions were calculated for multiple neutron energies, using the same code PROTON. The response function relates the neutron energy to the spectrometer measured quantity - proton energy.

	The  second topic focuses on application of the spectrometer in different neutral beam injection (NBI) heated DT plasma scenarios.   The spectrometer's ability to determine the DT fuel ion ratio  $n_t/n_d$ and fuel ion temperature (\textit{T})  was simulated using the SPEC code together with the previously calculated TPR spectrometer response functions. 
	
	\section{Neutrons for DT Plasma Diagnostics}
	\label{sec:DT_diagnostics}
	
	Normally, most of the fusion reactions occur inside the plasma core from where the neutrons are emitted. As a result, information determined by neutron plasma diagnostic techniques are often weighted towards the plasma core and can provide additional information to other diagnostic techniques. It is possible to determine the Deuterium and Tritium fuel ion ratio from the DD (2.5~MeV) and DT (14~MeV) neutron emission~\cite{Kallne1991}. However, in some DT plasma scenarios the DD neutron emission is indistinguishable from the neutron background, which is dominated by DT neutron back-scattering~\cite{Kallne1991,Ericsson2010} from the vessel wall. A possible solution to the issue is to use only the DT neutron emission for diagnostics~\cite{Hellesen2015,Hellesen2012}, which is the only method investigated in this paper.
	
	\begin{figure}[h]
		\centering
		\includegraphics[width=0.5\textwidth]{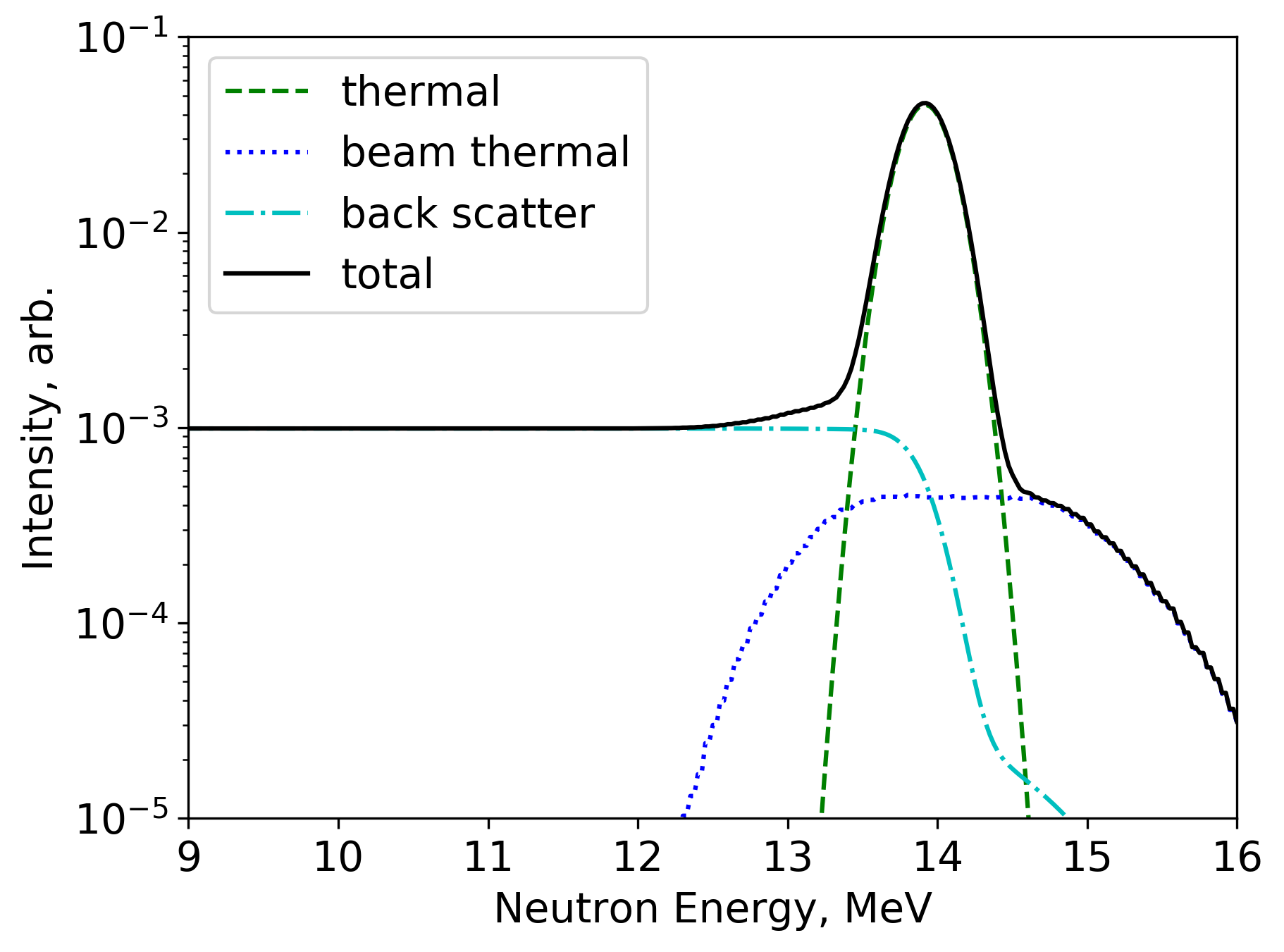}
		\caption{ Example of a synthetic neutron spectrum for a 10~keV DT plasma with back-scattering  (25\%) and beam-thermal power  (5\%) components. The beam-beam component is not significant in this example.}
		\label{fig:mymodel}
	\end{figure} 	
	
	In a NBI heated plasma the fuel ions can be separated into two populations: thermalised ions and fast beam ions. The two populations lead to three possible components dominating the neutron spectra: thermal, beam-thermal and beam-beam. The thermal component arises due to fusion reactions between thermalised fuel ions. The beam-thermal component arises due to fusion reactions between beam ions and thermalised ions, while beam-beam component originates from fast beam ion fusion. In this paper the latter component is expected to be small and is neglected. In addition to these components a back-scattering component will always be present in experimental neutron spectra. An example of such a  neutron spectrum containing thermal, beam-thermal and a back-scattering components is shown in Figure \ref{fig:mymodel}. The plasma temperature determination can be obtained from the energy spread  of the thermal component alone as detailed in \cite{brysk1973}. However, the  $n_t/n_d$  determination from the DT spectrum  is more complicated as it involves accurate modelling of the NBI slowing down distribution to determine the shape of the beam-thermal component~\cite{Hellesen2014}. The method in consideration is mostly applicable to spectra dominated by the DT thermal component. Assuming a deuterium beam, the neutron intensity from thermal, $I_{TH}$, and beam-thermal, $I_{NB}$, components can be expressed:
	\begin{equation}
	I_{TH} = {n_d n_t } \langle \sigma \nu \rangle_{TH}
	\label{eq:beamTH}
	\end{equation}
	\begin{equation}
	I_{NB} = {n_{NB} n_t } \langle \sigma \nu \rangle_{NB},
	\label{eq:beamNB}
	\end{equation}
	where $n$ is the density of the specified ion species, and $\langle \sigma \nu \rangle$ is the average reactivity for the interaction of the two ion species involved. The fuel ion ratio can then be expressed as:
	\begin{equation}
	\frac{n_t }{ n_d}  =  \frac  {I_{NB}^2 \langle \sigma \nu \rangle_{TH}} {I_{TH}  (n _{NB}\langle \sigma \nu \rangle_{NB})^2}
	\label{eq:ratio}
	\end{equation}	
	The main uncertainty contribution to the $n_t/n_d$ is due to $I_{NB}$ and, as  explained in \cite{Hellesen2014}, a 10\% uncertainty in $I_{NB}$ leads to  at least a 20\% uncertainty in the $n_t/n_d$ ratio determination.

	\section{Thin-foil Proton Recoil Spectrometer }
	\label{sec:TPR}
\begin{figure}[htbp]\centering
		\includegraphics[width=0.6\textwidth]{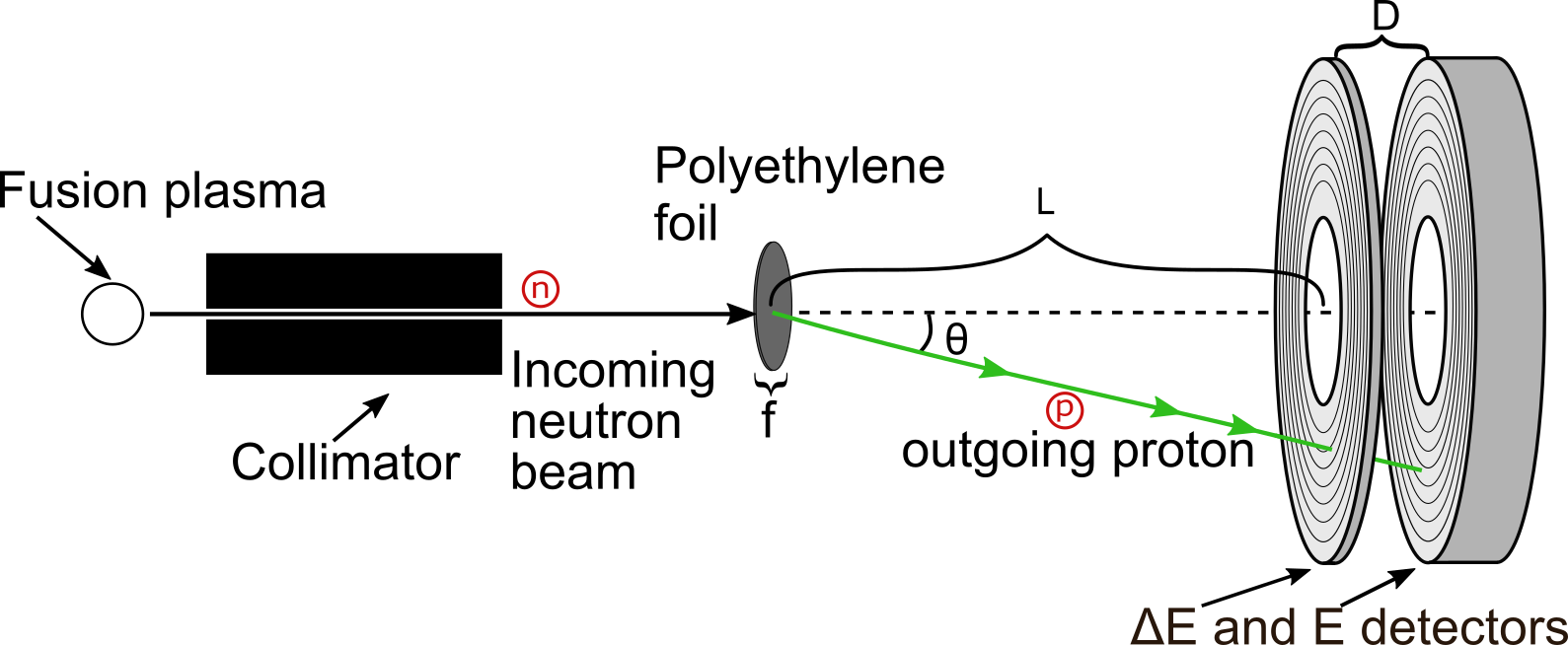}
		\caption{ A simplified scheme of a thin foil proton recoil spectrometer. $f$ - foil thickness,  \textit{D }- distance between detectors,  \textit{L} - distance from foil to $\Delta$E detector, $\theta $ is recoil proton  scattering angle. }
		\label{fig:geo}
	\end{figure}

	The TPR spectrometer considered here consists of a thin polyethylene foil as a neutron-to-proton converter and two annular silicon detectors as depicted in Figure~\ref{fig:geo}.  Neutrons leaving the plasma pass through a collimator and impinge on the polyethylene foil. Some of the neutrons elastically scatter on hydrogen nuclei in the foil and transfer some of the energy to the recoil protons. The  proton energy relates to the incoming neutron energy: \mbox{ $E_p = E_n  cos^2 (\theta )$}, where $\theta $ is the recoil proton scattering angle in the labratory system. Depending on $\theta $, the protons may hit an annular telescope detector consisting of two detectors in a $\Delta $E and E configuration. The neutron energy spectrum and plasma parameters can then be inferred from the measured proton energy spectrum.
	
	The two detectors in consideration are readily available segmented annular silicon detectors based on the Micron~\cite{micron} S1 double sided silicon detector design. The model considered has detectors with a radial division of 16 segments and thickness of 500 and 1500~$\rm{\mu m}$. The detector's inner and outer radii are 24 and 48~mm, respectively. The use of two detectors allows for coincidence measurements which can be used in the data analysis to improve the signal to background  ratio. In this paper we considered two types of detectors: segmented, i.e. taking advantage of the radial segments, and non-segmented, i.e. summing all radial segments. Silicon material is a good choice for the  $ \Delta $E detector due to good energy resolution at thicknesses below  500~$\rm{\mu m}$, which would be difficult to achieve using, for example, scintillation detectors.
	
	The spectrometer neutron energy range is mainly limited by the thicknesses of the silicon detectors. The total thickness of the mentioned detectors is sufficient to fully stop 18~MeV protons while at least 8~MeV is necessary to penetrate the first detector, thus in coincidence mode it can be expected to measure a proton energy range of about 8.5 to 18~MeV. Then, depending on the foil thickness and scattering angle, this would correspond to approximately 9 - 18~MeV neutron energy range.

	\section{Efficiency and Energy Resolution Calculations}
	\label{sec:eff_calculation}
	The spectrometer performance is affected by multiple geometrical parameters: foil thickness, foil area, detector thickness, distance between detectors and distance from first detector to the foil as well as detector geometry.  An increase of foil thickness, $f$, would  improve the efficiency due to increased interaction probability and degrade the energy resolution. For this example, the resolution can be affected by three factors: the interaction position uncertainty and proton energy loss in the foil as well as the proton straggling. Unfortunately the efficiency and energy resolution have an inverse correlation which requires optimisation of the spectrometer.
	
	We have selected to investigate three geometrical parameters: distance between foil and the first detector, $L$, foil thickness, $f$, and distance between the $\Delta $E and E detectors, $D$,  as shown in Figure~\ref{fig:geo}. The foil area was fixed at 10~$mm^2$, while the detectors in use are described in previous section. Each spectrometer configuration has an unique set of $f $, $L$ and $D$. A Python 3 based Monte-Carlo  code, PROTON, was developed to determine the efficiency and energy resolution by varying the selected parameters. The best spectrometer configurations were found using Pareto optimisation~\cite{2004Optimization}: all investigated  configurations were sorted according to their energy resolution and the configuration with the highest efficiency for a specific resolution was found.
	
	The code simulates a perpendicular 14~MeV neutron beam impinging on the polyethylene foil, then calculates proton energy and the angle after n,p scattering in the foil. PROTON also estimates the subsequent proton energy deposition in the foil and Si detectors. The angular straggling was estimated by adding an energy dependent Gaussian spread on the proton angle after the first Si detector. The expected angular spread was estimated using  SRIM~\cite{ZIEGLER1998} for multiple proton energies. Dedicated libraries were used for proton stopping power~\cite{pstar} and neutron elastic cross section \cite{ENDF} in silicon and polyethylene. After running this processes for a number of histories, a proton energy deposition distribution in the detectors is constructed and the full width at half maximum~(FWHM) is calculated.  
	
	The energy resolution  was calculated for two detector types: segmented and non-segmented. For the segmented detectors, any combination of segments from two different detectors acts as a detector.  The protons scattering towards the detector can have a scattering angle from $\theta_{min} $ to $ \theta_{max} $. For example, if we consider only protons interacting in the i-th segments in both of the detectors, the angle interval is limited by the inner   $ \theta_{i} $ to outer $ \theta_{i+1}  $ angles of the segment edges respectively, where $ \theta_{min} \leq \theta_i  \le \theta_{i+1}  \leq  \theta_{max}$. The resulting  distribution is a subset of the whole angular distribution. This leads to a smaller variance of the scattering angle compared to the mean and, as a result, improves the energy resolution. Hence, using the segmented detector type allows to put the detectors closer to the foil without degrading energy resolution and thereby increasing the efficiency, as compared to the non segmented. As a result the  FWHM calculation of the segmented detector was done in the following way: each detector segment proton energy deposition distribution was centred around  14~MeV and then all segment distributions were summed to create one energy deposition distribution. Then the FWHM of the created distribution was found. For the non segmented detector the FWHM was calculated directly from the obtained proton energy deposition distribution.
	
	PROTON also produces a spectrometer response function, $\overline {\overline {SRF}}$. The spectrometer response function is created for neutron energies varying from 9 to 18~MeV in 40~keV steps. Each sampled neutron energy creates a proton energy distribution, thus a $\overline {\overline {SRF}}$  column defines the proton energy probability distribution created by the neutrons impinging on the TPR spectrometer. Each $\overline {\overline {SRF}}$ column has a specific uniformly distributed neutron energy range. Using the discrete Fredholm equation the measured proton spectrum, $\overline E_p$, relates to a neutron spectrum, $\overline E_n$, by:

	\begin{equation}
	\overline E_p  = \overline{\overline {SRF}}  \cdot \overline E_n
	\end{equation}

	\section{Evaluation method}
	\label{sec:evaluation}
\begin{figure}[h]
		\centering
		\includegraphics[width=0.95\textwidth]{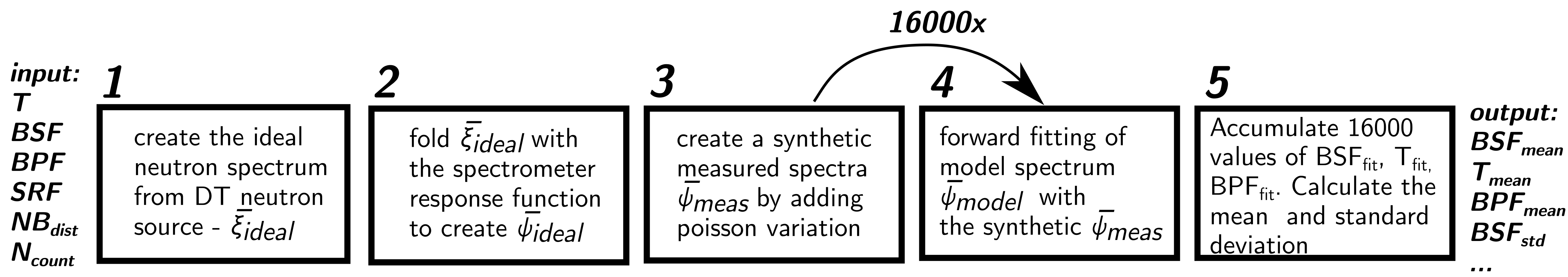}
		\caption{ SPEC calculation scheme. See text for details. }
		\label{fig:schema}
	\end{figure}

	The evaluation method applied in this paper is nearly identical to that in \cite{Sunden2013} using a Fortran code SPEC. The code takes the following input parameters: NBI power fraction (\textit{BPF}), thermal ion temperature (\textit{T}), back-scattering fraction (\textit{BSF}), neutral beam slowing down distribution ($NB_{dist}$), $\overline {\overline {SRF}}$ and an expected number of counts in the detector ($N_{{count}}$). These inputs are used to create a synthetic ideal and measured neutron energy spectrum. The output provides the accuracy and precision of the neutron spectrometer in measuring \textit{BSF}, \textit{BPF} and \textit{T} (see  Figure~\ref{fig:schema}).
	Both the $BPF$ and $BSF$  component intensities are defined as a fraction of total plasma neutron emission intensity. The same neutral beam slowing down distribution was used in all simulations and it had previously been calculated for a 500~MW plasma scenario in ITER conditions \cite{velocity}.
	
	The SPEC code computation can  roughly be  divided into five stages, as shown in Figure~\ref{fig:schema}. First an ideal synthetic neutron spectrum,  $\overline{ \xi} _{ideal}$, is constructed for each set of  \textit{T}, \textit{BSF}, \textit{BPF} and $NB_{{dist}}$. A spectrum example is shown in Figure~\ref{fig:mymodel}. During the second step, $ \overline{ \xi} _{ideal}  $ is folded with the spectrometer response function, $ \overline {\overline {SRF}}$, for a specific plasma scenario which results in a spectrometer response. The spectrometer response is  normalised to the expected number of counts as shown in eq.~\ref{eq:model} to produce $\overline{ \psi} _{ideal}  $.  The calculated $\overline{ \psi} _{ideal}  $ is an ideal binned proton energy distribution, for the TPR spectrometer, given by the synthetic neutron spectrum $\xi_{ideal} $:
	
	\begin{equation}
	\label{eq:model}
	{\psi} _ {ideal,i} = N_{count} \cdot \frac {   {SRF}_{i,j} \times  {\xi} _{ideal,j}  }{ \sum_{i=0}^{max}  {SRF}_{i,j} \times  {\xi} _{ideal,j} } 
	\end{equation}
	
\begin{figure}[htbp]
		\centering
		\includegraphics[width=0.44\textwidth,origin=c,angle=0]{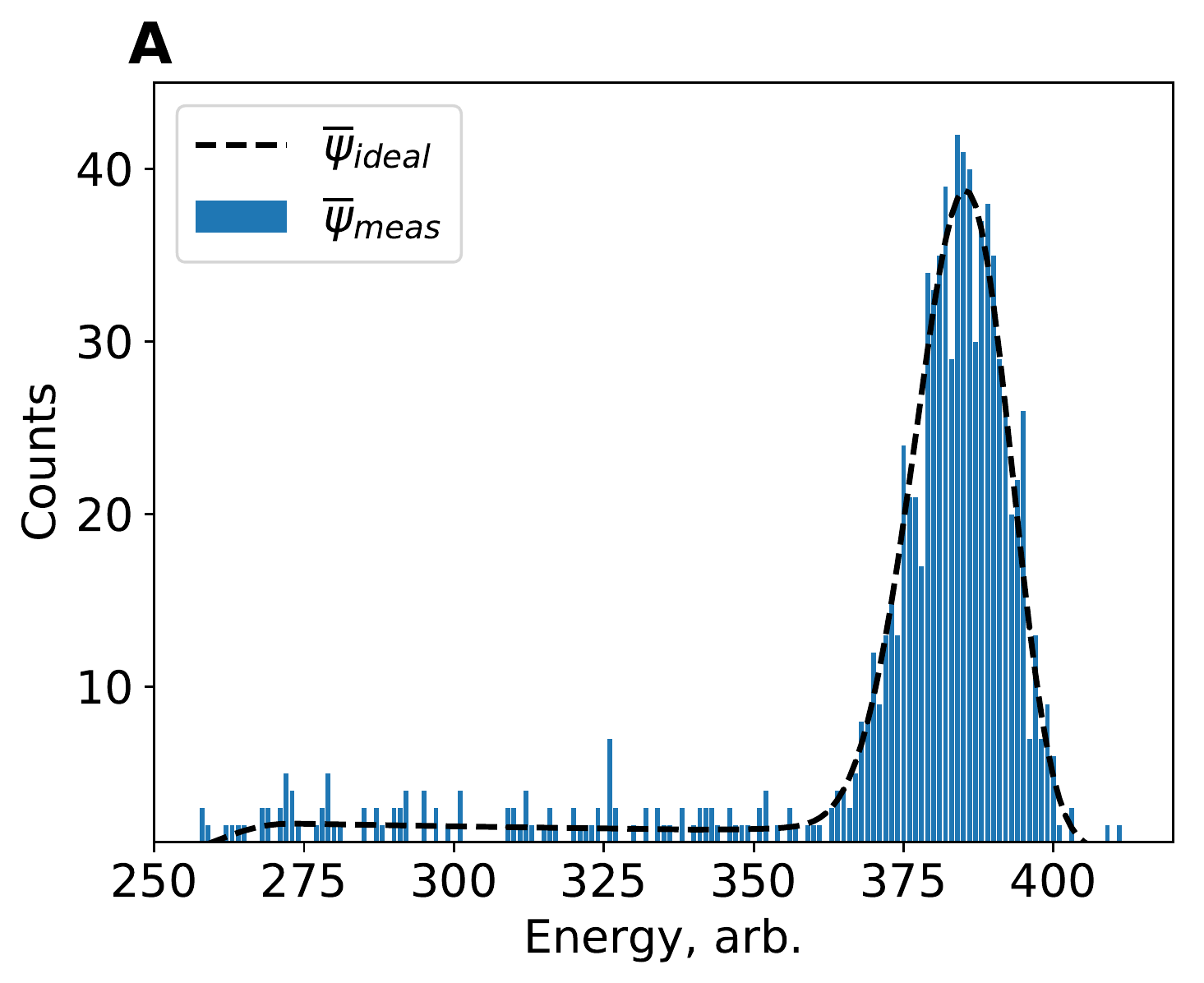}
		\qquad
		\includegraphics[width=0.45\textwidth,origin=c,angle=0]{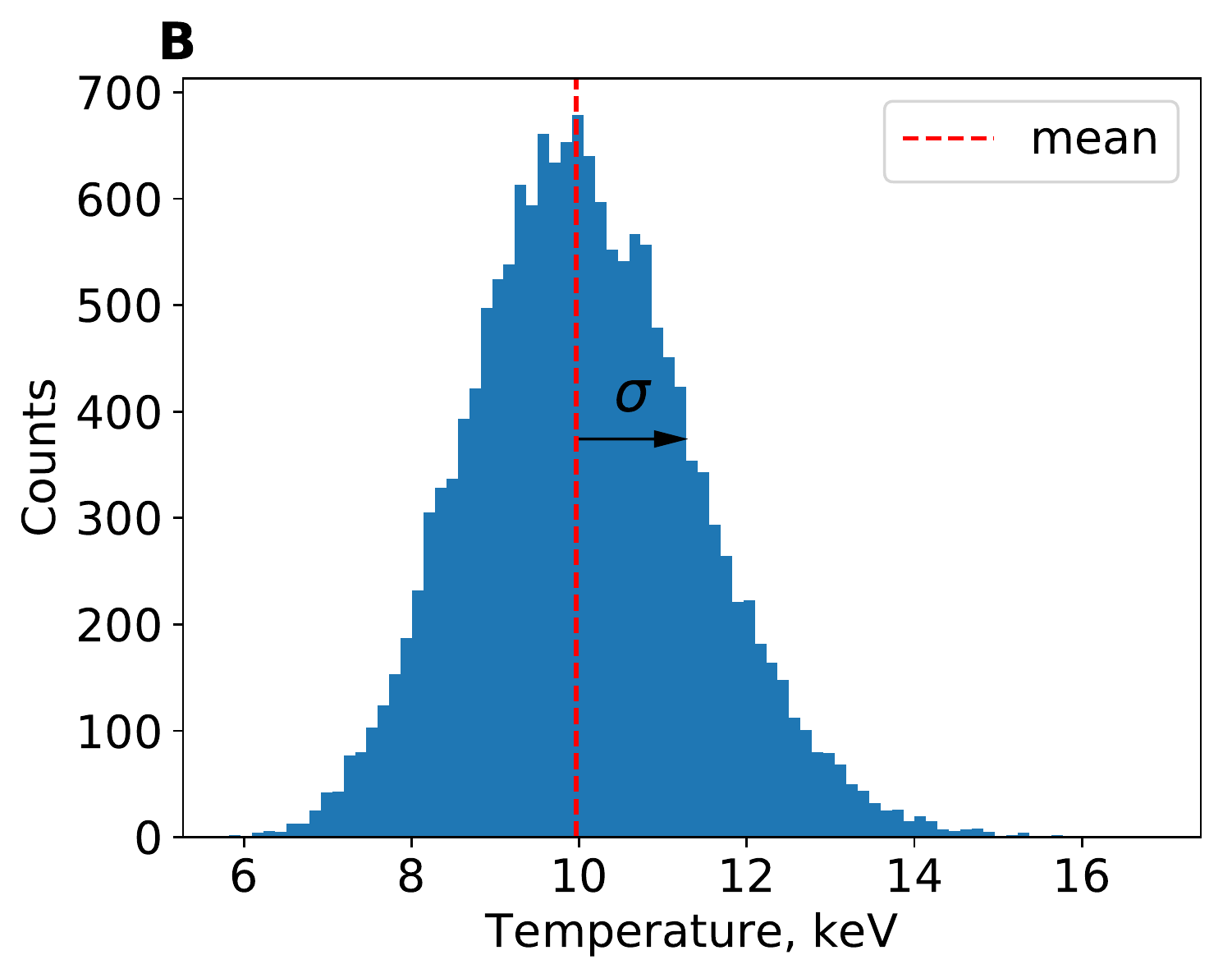}
		\caption{ A - Example of one synthetic measured spectrum  $\overline{ \psi} _{meas}  $   with a corresponding ideal  spectrum $\overline{ \psi} _{ideal}$ (dashed line). B -  Distribution of the plasma temperature values for a SPEC run with 16000 iterations. The mean is marked by dashed vertical line. The deviation of the distribution defines the spectrometer precision. }
		\label{fig:folded}
	\end{figure}

	During the third step we create a synthetic measurement spectrum, $\overline {\psi}_{meas}$. We use $ \psi_{ideal,i} $  as  mean value for a Poisson distribution and for each measurement bin, $\psi_{meas,i}$, a value is drawn from the corresponding Poisson distribution $ \psi_{meas,i} = Poiss(\psi_{ideal,i}  ) $. An example of one such synthetic measured spectra for a specific plasma scenario is shown in Figure~\ref{fig:folded}A, where \textit{T}~=~10~keV, \textit{BPF}~=~0.05, \textit{BSF}~=~0.25 and $N_{count}$~=~1000.
	
	The fourth step is to perform a forward convolution fitting to find the best estimates of the plasma parameters describing  $ \overline{ \psi} _{meas} $. A plasma parameter space is sampled with different \textit{T}, \textit{BPF}, and \textit{BSF} values within a reasonable user specified range. Then an ideal spectrum is made for each combination of the plasma parameters as in step 1 of Figure~\ref{fig:schema} and each of the ideal spectra are convoluted with the spectrometer response function to produce  $ \overline{\psi} _{model}(T,BPF,BSF) $ as in step two.  For each $ {\psi}_{meas}$  we calculate the modified Cash statistic C value~\cite{cash1979,Arnaud1996}, where the minimum C value corresponds to the most likely $ {\psi}_{model}$ and gives us the   combination of \textit{T}, \textit{BPF}, \textit{BSF}:
	\begin{equation}
	C =   2 \cdot \sum_{i=0}^{max}(  {\psi}_{model,i}  -  {\psi}_{meas,i} +  {\psi}_{meas,i} \cdot ln(  {\psi}_{meas,i} / {\psi}_{model,i} ) ) 
	\label{eq:cash}
	\end{equation}
	where i represents measurement bin.
	
	Steps 3 and 4 are repeated 16000 times (as shown in Figure~\ref{fig:schema}) until a distribution of estimated plasma parameters is created. The last step of the program is to calculate the mean and standard deviation of the fitted plasma parameter distributions.
	The difference between the mean of the distribution and the true value, used to produce $\xi_{ideal}$,  represents the accuracy that can be obtained from a measurement. An example distribution in measuring  10~keV temperature   is shown in Figure~\ref{fig:folded}B. A SPEC run was performed for each set of input temperature, beam fraction and detector count rate. During the same SPEC run one distribution for each parameter \textit{T}, \textit{BPF} and \textit{BSF} is created.  The \textit{BSF} was kept at  constant fraction of 0.25 for all runs. 
	
	\section{Results and Discussion}
	\label{sec:results}
	\subsection{Resolution Simulations}
	
\begin{figure}[ht]
		\includegraphics[width=\textwidth]{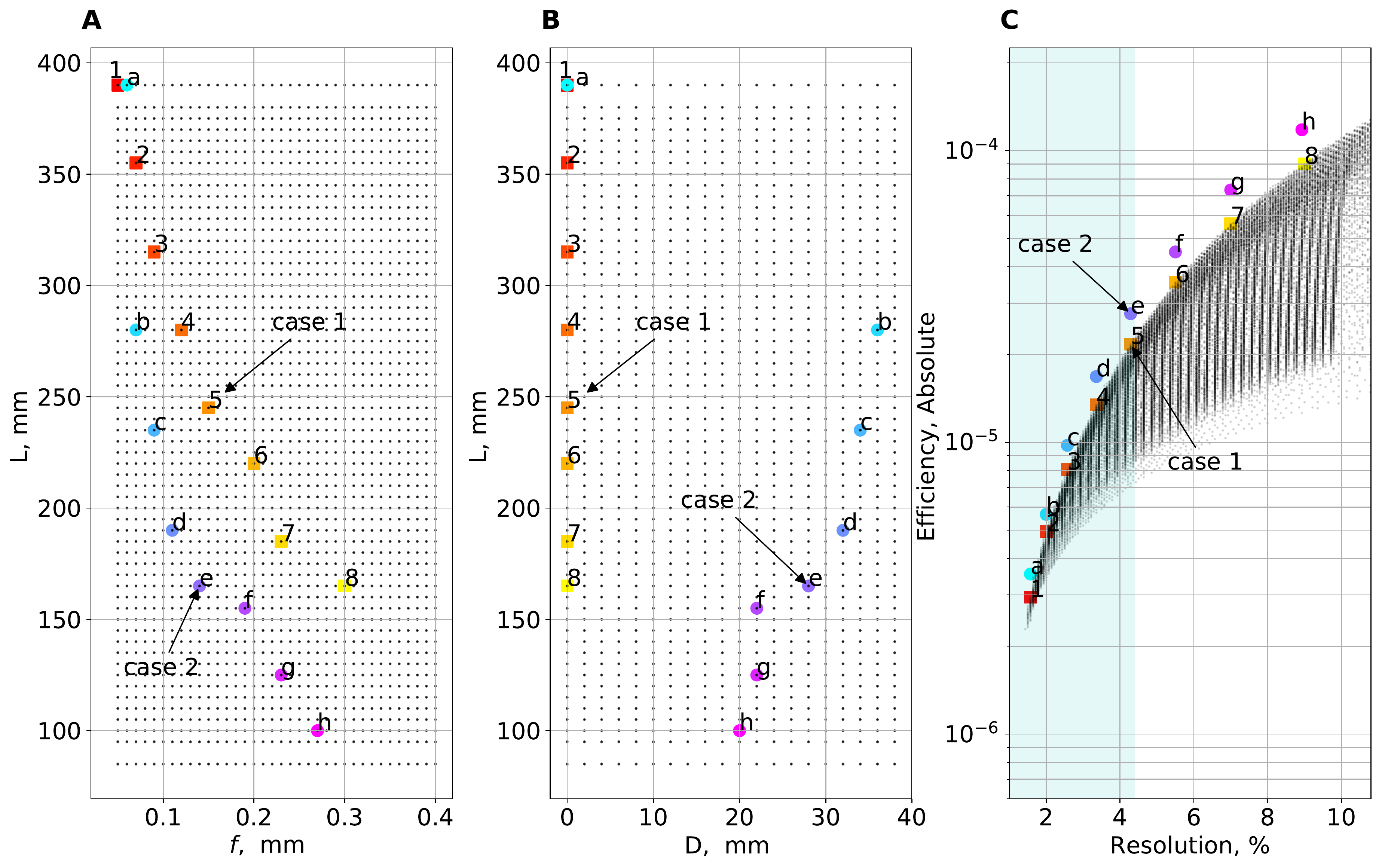}
		\caption{ Figure  A,B - investigated spectrometer geometrical parameter space \textit{L} -  distance between foil and first detector, \textit{f} - foil thickness and \textit{D} - distance between detectors, C - Pareto plot of efficiency vs. energy resolution. The resolution was calculated assuming that the detector is non-segmented (number labels) and segmented (letter labels). The case 1 and 2 are selected configuration examples. The points in all panels are interrelated.}
		\label{fig:effres}	
	\end{figure}

	In total 50400  spectrometer configurations were simulated using PROTON for both the segmented and  non-segmented spectrometer configurations. The configuration phase space is 3 dimensional with \textit{L}, \textit{f} and \textit{D} as variables, as shown Figure \ref{fig:geo}. Figure~\ref{fig:effres}A and B  are 2D projections of the phase space and each black dot represents an investigated configuration.  For each spectrometer configuration PROTON generated  $2 \cdot 10^6$ useful events to find a maximum efficiency for each energy resolution as shown in the Figure~\ref{fig:effres}C. All of  the labelled points are inter-related between the panels. For example the point labelled "e" corresponds to same spectrometer configuration in all of  Figure~\ref*{fig:effres} panels. The black dots in Figure~\ref{fig:effres}C  marks all non-segmented detector configurations and, as one can see, the upper edge , a.k.a the Pareto front, marks the highest efficiency designs. The  algorithm has selected configurations with the highest efficiency for a specific resolution, marked points are labelled by numbers for the non segmented detector; while the letter labelled points mark the Pareto front for the segmented detector configurations. Only the optimum non segmented detector designs are marked in Figure~\ref{fig:effres}C for readability reasons. \newline
	
	Looking at the numbered points of Figure \ref*{fig:effres}B, we see that for a non-segmented spectrometer configuration a small distance between the two detectors is favoured. For the segmented detector the distance between detectors varies depending on the energy resolution. For an energy resolution below \textasciitilde 2\% the gain of segmentation is insignificant due to proton angular straggling and as a result, small distance between detectors is favoured as seen from points (a) and (1) in Figure~\ref*{fig:effres}B and C. High efficiency configurations have the smallest \textit{L} to increase the solid angle covered by the detector. In general the segmentation results in a more compact configuration and efficiency improvement of at least 10\%, for the same energy resolution. However, the advantages are more visible for energy resolution above 2\%. In this paper we discuss the two selected configurations labelled case~1 and case~2 in Figure~\ref{fig:effres}C. Both of the configurations were selected to have a resolution of 4.3\%.  The  efficiency for case~2 is \textasciitilde 20\% higher than for case~1, as presented in Table~\ref*{tab:configuration}. 
	\begin{table}[htbp]
		\centering
		\smallskip
		\begin{tabular}{|lr|c|c|c|c|c|}
			\hline Configuration  &  \textit{D}, mm &  \textit{L}, mm  & \textit{ f}, mm &   efficiency  & FWHM \\ 
			\hline Case 1 & 0 & 245 & 0.15 & $\rm2.17 \cdot 10^{-5}$ & 4.3\%\\ 
			Case 2 & 28 & 165 & 0.14  & $ \rm2.76 \cdot 10^{-5}$& 4.3\% \\ 
			\hline 
		\end{tabular} 
		\caption{Selected configurations for segmented and non-segmented detectors. The \textit{D},\textit{ L }and \textit{f} correspond to the Figure \ref{fig:geo}. }	
		\label{tab:configuration}
	\end{table}	
	
\begin{figure}[htbp]
		\centering
		\includegraphics[width=.43\textwidth,angle=0]{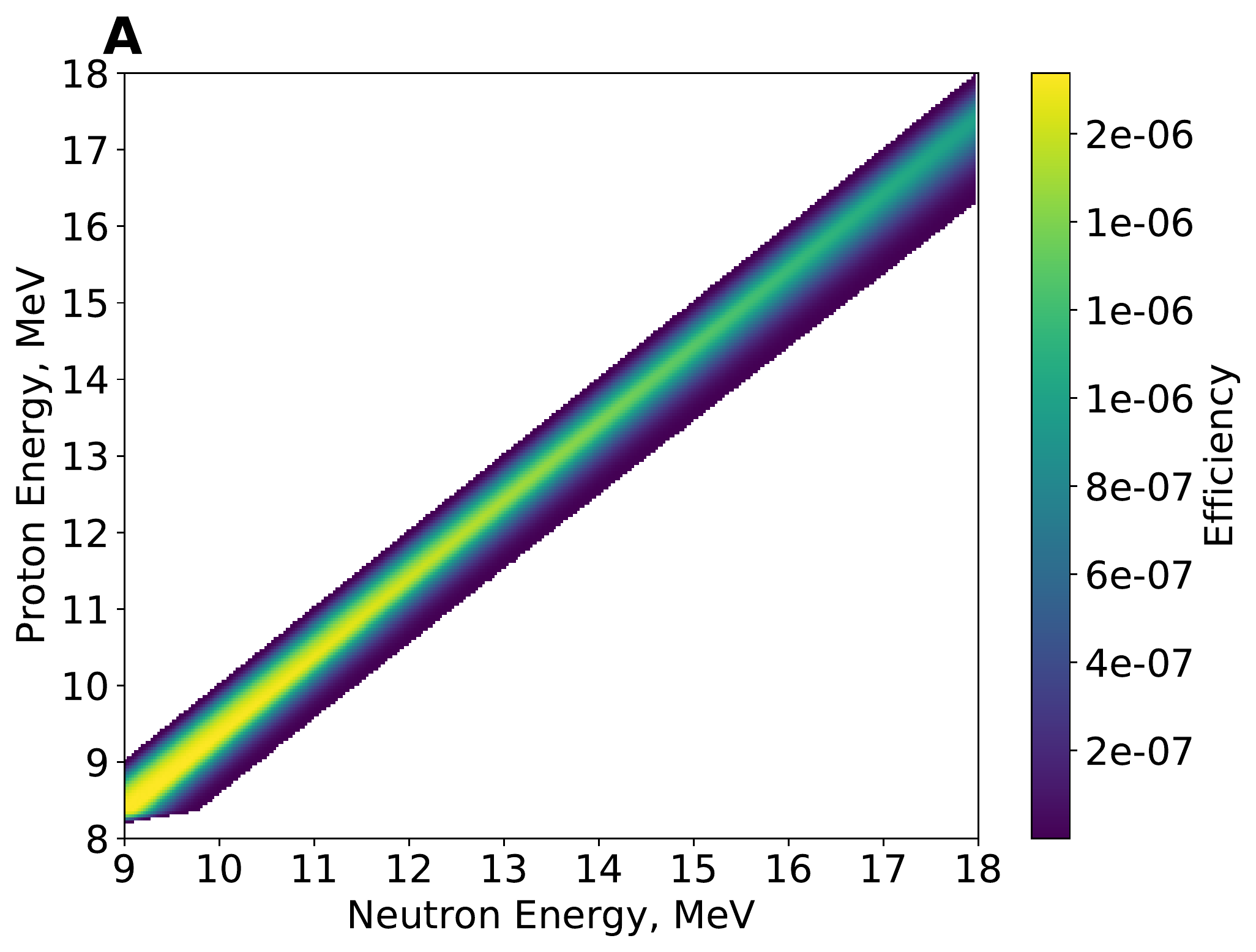}
		\qquad
		\includegraphics[width=.48\textwidth,angle=0]{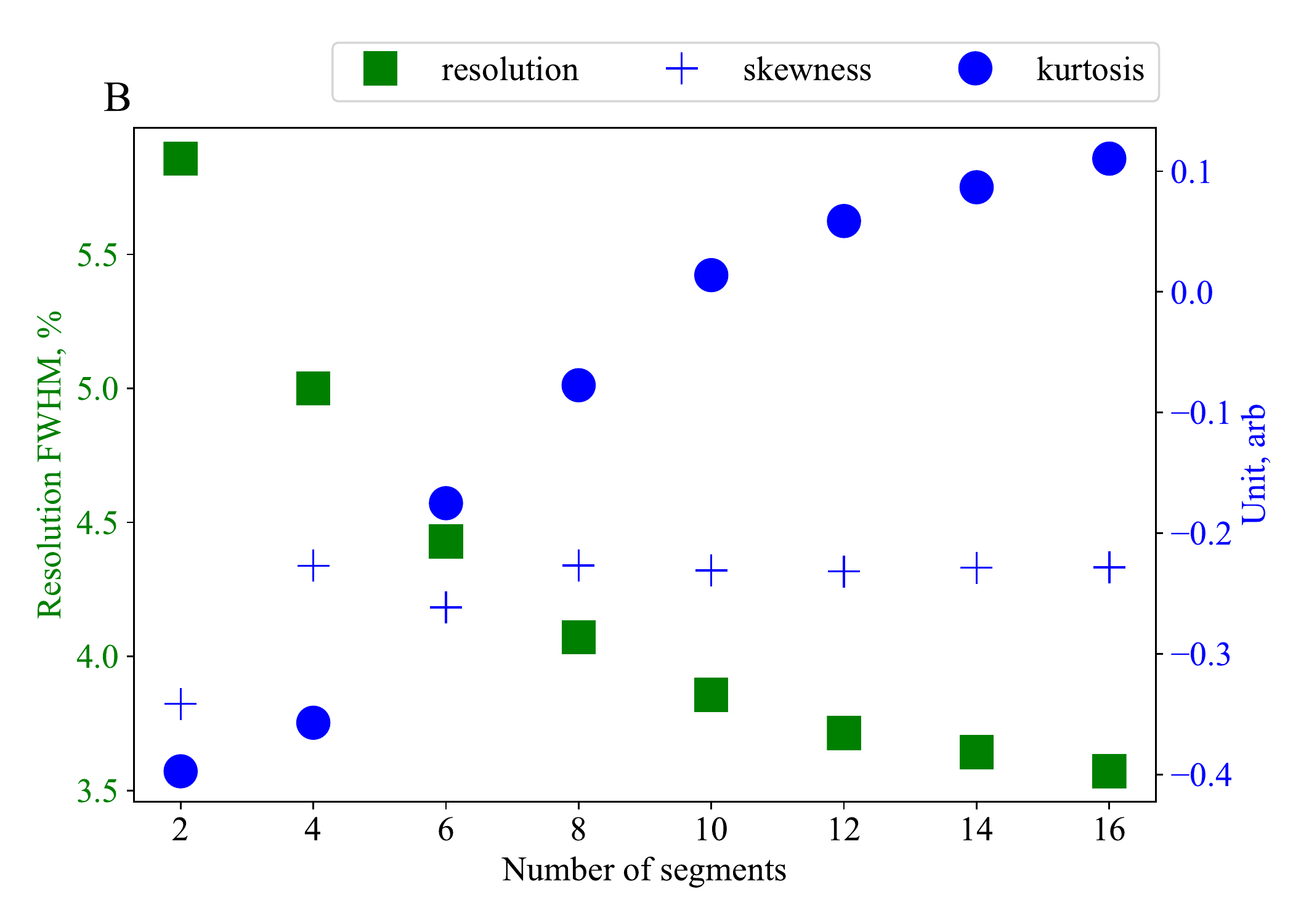}
		\caption{ A -  Case 1 spectrometer response function used in the plasma parameter estimation. 
			B -  proton distribution shape and resolution dependence on the detector radial segmentation. The left green y-axis marks resolution while the right  blue axis marks unit values for kurtosis and skewness. 
		}
		\label{fig:response}
	\end{figure}

	Case~1~and~2  spectrometer configurations (shown in Figure~\ref{fig:effres}) were selected and PROTON calculated  the corresponding spectrometer response functions; the case~1 response function is shown in Figure~\ref{fig:response}A. The x-axis marks incoming neutron energy  while the y-axis and colour coding gives the proton energy probability distribution for the selected neutron energy. Total proton energy is calculated by summing the energy deposited in both $\Delta $E and E detectors.

	We have calculated the resolution and proton energy distribution skewness and kurtosis dependence on the amount of detector segments as show in Figure~\ref{fig:response}B. The presented comparison is for the spectrometer case~2 configuration. In combination with the figure and currently selected 16 segment detector we have chosen a division to 8 segments. This is straightforward to implement and gives sufficient energy resolution without increasing the system cost significantly. The resolution does not improve linearly with increasing segment count due to the angular straggling of the proton beam after the first detector. In addition in Figure~\ref{fig:response}B it can be seen that skewness and kurtosis are close to 0, which is expected of a  Gaussian distribution, and we can assume that the spectrometer response function is close to Gaussian. All previous calculations assumed that the segmented detector had 8 segments. It should be kept in mind that detectors intrinsic resolution influence on the energy resolution is not estimated, thus it is expected that the spectrometer performance in experimental conditions will be somewhat worse, however this should be a fairly small effect.

	\subsection{Spectrometer Performance}
	The spectrometer's performance was evaluated  using SPEC and TPR spectrometer response functions. We have estimated the spectrometer's ability  to determine  $n_t/n_d$  and $T$ as well as the expected precision. The spectrometer's performance was evaluated by varying the fusion plasma parameters in the following way:  \textit{T} was varied from 4 to 20 keV with step length of 2~keV, \textit{BPF} from 0.03 to 0.25 with a step length of 0.01, and the \textit{BSF} was fixed at 0.25. The $N_{count}$ (equation \ref{eq:model}) was evaluated between 600 to 38000 counts in 53 steps of increasing length. The selected  parameter range were chosen to cover a wide range of DT plasma types of interest for ITER.
	
	\begin{figure}[htbp]
		\centering
		\includegraphics[width=.98\textwidth]{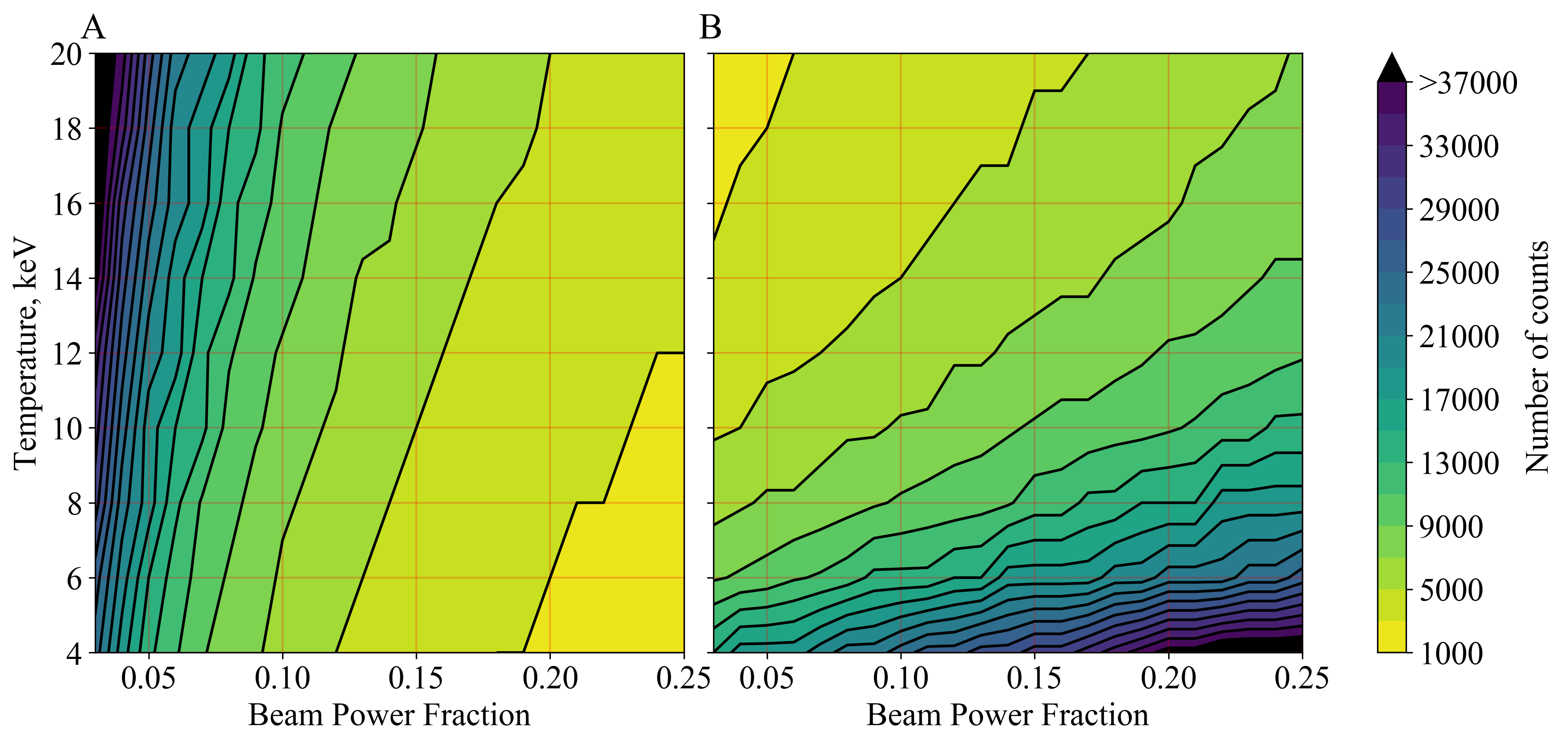}
		\caption{ Colour map of spectrometer counts required to achieve 20\% precision in fuel ion ratio determination and 10\% precision in temperature determination as a function of temperature and beam power fraction for panels A and B respectively.}
		\label{fig:precision}
	\end{figure}
	
	The minimum required counts $N_{min}$ to determine $n_t/n_d$ with 20\% and \textit{T}  with 10\% precision  as a function of temperature and beam power fraction are shown in Figure~\ref{fig:precision} panels A and B, respectively. As seen from the figures the spectrometer performance is correlated with the \textit{BPF} and \textit{T}. The required measurement time \textit{t} for a specific \textit{BPF} and \textit{T} can be determined using: $  t  =  N_{min} \div (\epsilon \cdot I_n )$, assuming we know the neutron intensity \textit{$I_n$}, efficiency $\epsilon$ and number of counts required to achieve the precision $N_{min}$. For example, at ITER full power scenario with \textit{BPF} of 0.06 and \textit{T} of 10~keV  one expects a neutron intensity of  $10^{10} \ { \frac{n}{s} }$ at the detector foil. Taking the efficiency from  Table~\ref*{tab:configuration} and $N_{min}$ from Figure~\ref{fig:precision}A we can determine  $n_t/n_d$ with a precision of 20\% within a $\tau_{D,T} \approx$ 70~ms integration window. The same can be estimated for the temperature using Figure \ref*{fig:precision}B which leads to a  $\tau_{T}\approx $ 27~ms window. 
	
	The spectrometer response function for the case~2 configuration is larger which leads to more than an order of magnitude longer simulation time. For this reason only a subset of the 47 plasma scenarios were used and compared with the case~1  results. The case~2  results show a slightly better precision in measuring temperature assuming the same number of counts, no other statistically significant differences were observed.  The case~2 measurement times, for the previously discussed plasma conditions, are: $\tau_{D,T}  \approx$~55~ms and $\tau_{T}\approx $~21~ms, which is slightly better than for the case~1.
	
	\section{Conclusions and Outlook}
	The spectrometer simulations presented here demonstrate that taking advantage of the  detector segmentation could improve efficiency by at least 10\% and decrease the spectrometer size. The results on  fuel ion ratio and ion temperature determination provide a capability to estimate required integration time for different ITER-like plasma conditions, assuming the neutron intensity is known.  This result can be used for other spectrometers with Gaussian-like response functions for a rough estimate.
	
	Assuming a combination of two spectrometers and a neutron intensity of $10^{10} \ {\frac{n}{s} }$ the temperature and $n_d$/$n_t$ can be determined within ITER requirements of 100 ms integration window for any combination of \textit{BPF} ranging from 0.05 to 0.25 and T from 5 - 20~keV with a back-scattering fraction of 0.25.

	Future work should cover a more extensive analysis of the selected spectrometer configurations using dedicated Monte-Carlo codes to calculate the response functions, estimate background contribution as well as the spectrometer's ability to discriminate the background.  This would take into account  higher order effects not considered in this study. In addition there are plans to test a prototype spectrometer in coincidence mode using proton and DT neutron sources.


\end{document}